# LOCx2-130, a low-power, low-latency, 2 x 4.8-Gbps serializer ASIC for detector front-end readout

**Le Xiao[a,b], Quan Sun[b], Datao Gong[b], Emily Baker[c], Binwei Deng[d,b], Di Guo[a,b], Huiqin He[e], Suen Hou[f], Chonghan Liu[b], Tiankuan Liu[b,1], James Thomas[b], Jian Wang[b,g], Annie C. Xiang[b], Dongxu Yang[b,g], Jingbo Ye[b], Xiandong Zhao[b], Wei Zhou[a,b]**

[a] *Department of Physics, Central China Normal University, Wuhan, Hubei 430079, P.R. China*

[b] *Department of Physics, Southern Methodist University, Dallas, TX 75275, USA*

[c] *Keller High School, Keller, TX 76248, USA*

[d] *Hubei Polytechnic University, Huangshi, Hubei 435003, P.R. China*

[e] *Shenzhen Polytechnic, Shenzhen 518055, P.R. China*

[f] *Institute of Physics, Academia Sinica, Nangang 11529, Taipei, Taiwan*

[g] *State Key Laboratory of Particle Detection and Electronics, University of Science and Technology of China, Hefei Anhui 230026, P.R. China*

*E-mail*: `tliu@mail.smu.edu`

ABSTRACT: In this paper, we present the design and test results of LOCx2-130, a low-power, low-latency, dual-channel transmitter ASIC for detector front-end readout. LOCx2-130 has two channels of encoders and serializers, and each channel operates at 4.8 Gbps. LOCx2-130 can interface with three types of ADCs, an ASIC ADC and two COTS ADCs. LOCx2-130 is fabricated in a commercial 130-nm CMOS technology and is packaged in a 100-pin QFN package. LOCx2-130 consumes 440 mW and achieves a latency of less than 40.7 ns.

[1] Corresponding author.

## Contents

## 1. Introduction

The Large Hadron Collider (LHC) will be upgraded in 2018-2019 to reach about three times the current luminosity. In order to exploit fully the physics potential of the high luminosity, the phase-I upgrade of the ATLAS Liquid Argon (LAr) Calorimeter will be carried out during the second long shutdown period of LHC [1]. One of the main tasks of the ATLAS LAr Calorimeter is to upgrade the trigger system to suppress background noise. In the ATLAS LAr Calorimeter trigger readout system, about 34,000 analog channels need to be sampled, digitalized, and transmitted via 124 LAr Trigger Digitizer Boards (LTDBs) [2]. Each LTDB transmits digitized data out of the front end at the rate of about 200 Gigabit per second (Gbps). Optical links are widely used in LHC experiments due to their advantages of high bandwidth, high channel density, low mass, and no ground loop [3-4]. Therefore, a data transmission optical link has become the favored choice for the ATLAS LAr Calorimeter trigger upgrade.

The generic block diagram of an optical data transmission link is shown in Figure 1, including the transmitting part and the receiving part. The transmitting part is composed of an encoder, a serializer, and an optical transmitter. The function of the optical transmitter is to convert the electrical signal of high-speed serial data into the optical signal using a laser driver [5, 6] and a laser diode. The serializer is responsible for converting parallel data into high-speed serial data. On the receiving side, the optical receiver is responsible for recovering the electrical signal from the optical signal. The signal is then recovering to parallel data through a deserializer. The parallel data are encoded before transmission in the encoder. The encoder guarantees the DC balance of the serial data so that the clock-data-recovery (CDR) circuit of the receiver can recover a clock from the serial data. The encoder also inserts a frame header and/or trailer for data boundary identification and error detection. Accordingly, the decoder of the receiver is responsible for recovering the original data and checking whether an error has occurred during the data transmission.

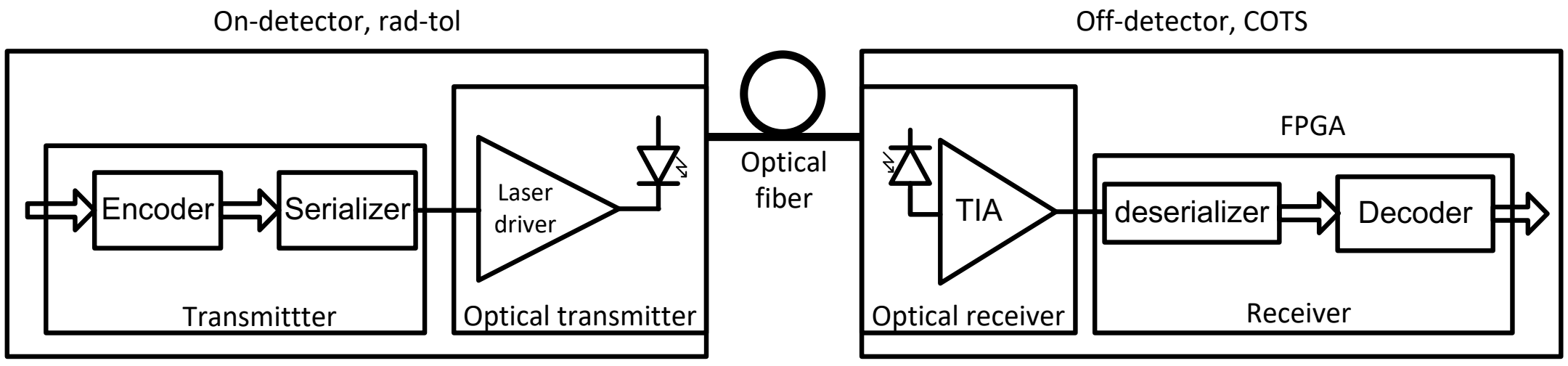

**Figure 1:** Block diagram of optical data transmission system.

The components on the receiving side can be implemented with Commercial Off-The-Shelf (COTS) components. However, the components on the transmitting side operate in a high-intensity radiation environment, and it is necessary to meet the radiation tolerance standard of ATLAS electronics [7-9]. In addition, as part of the trigger system, the latency of the entire optical fiber link must be less than 150 ns (not including the propagation time through the optical fiber), and the power consumption requirement of data transmitter is less than 100 mW per Gbps (not including the optical transmitter) [1].

In this paper, a transmitter AISC, LOCx2-130 is presented. LOCx2-130 is designed as a pin-compatible backup of LOCx2 [10], the baseline design in the ATLAS LAr Calorimeter Phase-I trigger upgrade. LOCx2-130 is a two-channel transmitter ASIC and each channel outputs serial data at the rate of 4.8 Gbps. LOCx2-130 is fabricated with a commercial 130-nm CMOS technology. LOCx2-130 meets all the link requirements with half of the power consumption of LOCx2.

The remainder of the paper is organized as follows: Section 2 describes the input and output interfaces. The design of LOCx2-130 is presented in Section 3. Section 4 discusses the LOCx2-130 test setup and measurement results. Section 5 summarizes the paper.

## 2. Input and output interfaces of LOCx2-130

The input data of each encoder come from either two ASIC ADCs called Nevis ADCs [11] or one of two COTS ADCs, ADS5272 [12] and ADS5294. It should be noted that the different types of ADCs cannot be used simultaneously. Each Nevis ADC has four analog channels, whereas each ADS5272 or ADS5294 has 8 analog channels. All ADCs sample analog signals at 40 MSamples/s. The resolution of the Nevis ADC is 12 bits ($D_{11}$-$D_0$). However, the output data of each analog channel in every conversion contains two extra bits for calibration data ($D_{13}$-$D_{12}$) and 2-bits of dummy data ($D_{15}$-$D_{14}$). The converted data from each analog channel of the ADC are serialized and then output at 640 Mbps. Each Nevis ADC provides a data clock (SCK) of 320 MHz to acquire the serial data and a frame clock (FCK) of 40 MHz to indicate the boundary of each sample. The resolutions are 12 bits ($D_{11}$-$D_0$) and 14 bits ($D_{13}$-$D_0$) and the output data of each analog channel are serialized and sent out at data rates of 480 Mbps and 560 Mbps for ADS5272 and ADS25294, respectively. Each ADS5272 or ADS5294 provides a data clock (SCK) of 240 MHz (ADS5272) or 280 MHz (ADS5294) and a frame clock (FCK) of 40 MHz.

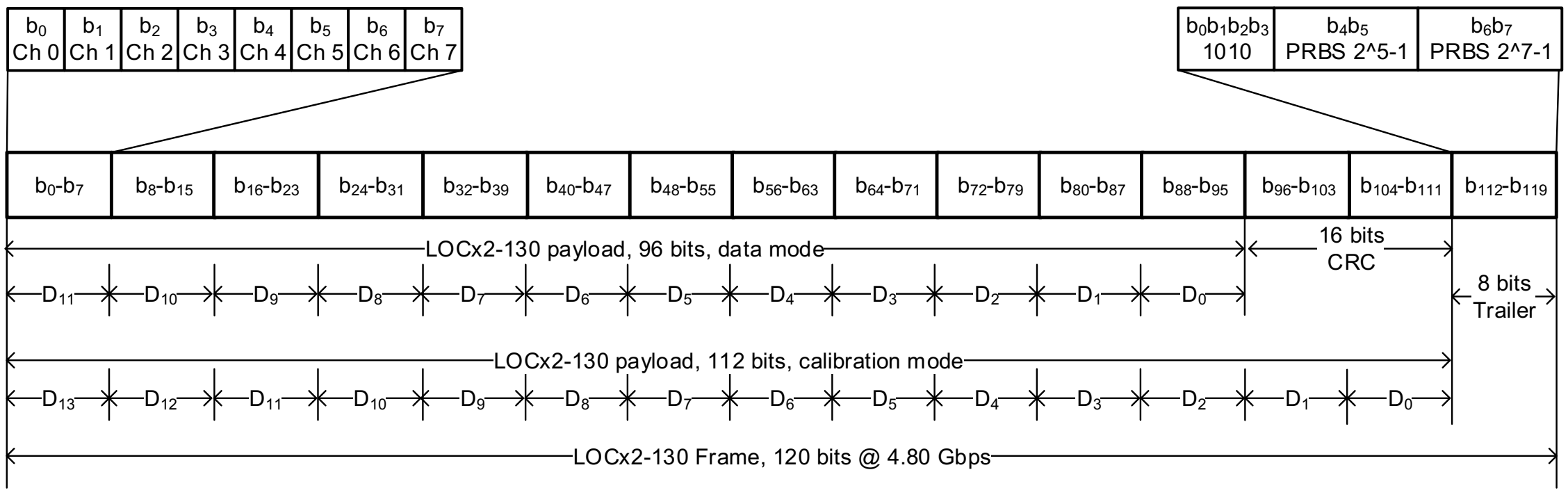

**Figure 2:** LOCic frame definition.

A custom line code is implemented in LOCx2-130 to prepare the data of the ADCs for each serializer. The 120-bit data that is transmitted each LHC clock cycle is defined as a frame. Figure 2 shows the frame definition of LOCx2-130. LOCx2-130 has two operation modes, the data mode and the calibration mode. In the data mode, each frame includes a 96-bit payload, a 16-bit Cyclic Redundant Check (CRC), and an 8-bit frame trailer. The data mode of LOCx2-130 is used for ADS5272 and Nevis ADCs when 2-bit calibration information is discarded. In the calibration mode, each frame includes a 112-bit payload and an 8-bit frame trailer. The calibration mode is used for ADS5294 and Nevis ADCs when 2-bit calibration information is kept.

The payload includes the data of eight ADC channels, represented as analog Channels 0-7 in Figure 2. Each ADC sample has 12 or 14 bits, represented as $D_{11}$-$D_0$ or $D_{13}$-$D_0$. Channel 0 is transmitted first and Channel 7 is transmitted last. The payload is scrambled to keep the signal DC-balanced in serial data transmission, whereas neither the CRC nor the frame trailer is scrambled.

There is a 16-bit CRC code in the data mode. There is no CRC protection in the calibration mode, which allows for the 14-bit data from each ADC channel to be transmitted in each conversion period. The CRC code is calculated from the unscrambled payload and is used to detect transmission errors.

The first four bits of the frame trailer is a fixed code 1010 used as the frame boundary. The other four bits come from two groups of 2-bit Pseudo-Random Binary Sequences (PRBSs) and forms a Bunch Cross IDentification (BCID) field. By combining the BCID fields of four consecutive frames, we can recover corresponding BCID information.

## 3. Design of LOCx2-130

The block diagram and the interface with other circuits of LOCx2-130 are shown in Figure 3. Each transmitter channel of LOCx2-130 comprises a LOCic-130 encoding unit, a 30:1 serializer, and an output driver. The two transmitter channels share a Phase-Locked Loop (PLL) and an Inter-Integrated Circuit ($I^2C$) slave. Only the interface with Nevis ADCs is shown in the figure. The encoding unit LOCic-130 prepares the ADC data for the following 30:1 serializer. Through the $I^2C$ slave, the GBT-based control link [13] can configure the internal registers of LOCx2-130. The GBT-based control link also supplies a 40 MHz LHC reference clock and a BCID reset for LOCx2-130. The serialized data are sent to an optical module named MTx [14]. LOCx2-130 is installed underneath an MTx on the LTDB board. In other words, LOCx2-130 and MTx are close to each other. Therefore, no pre-emphasis is used in LOCx2-130.

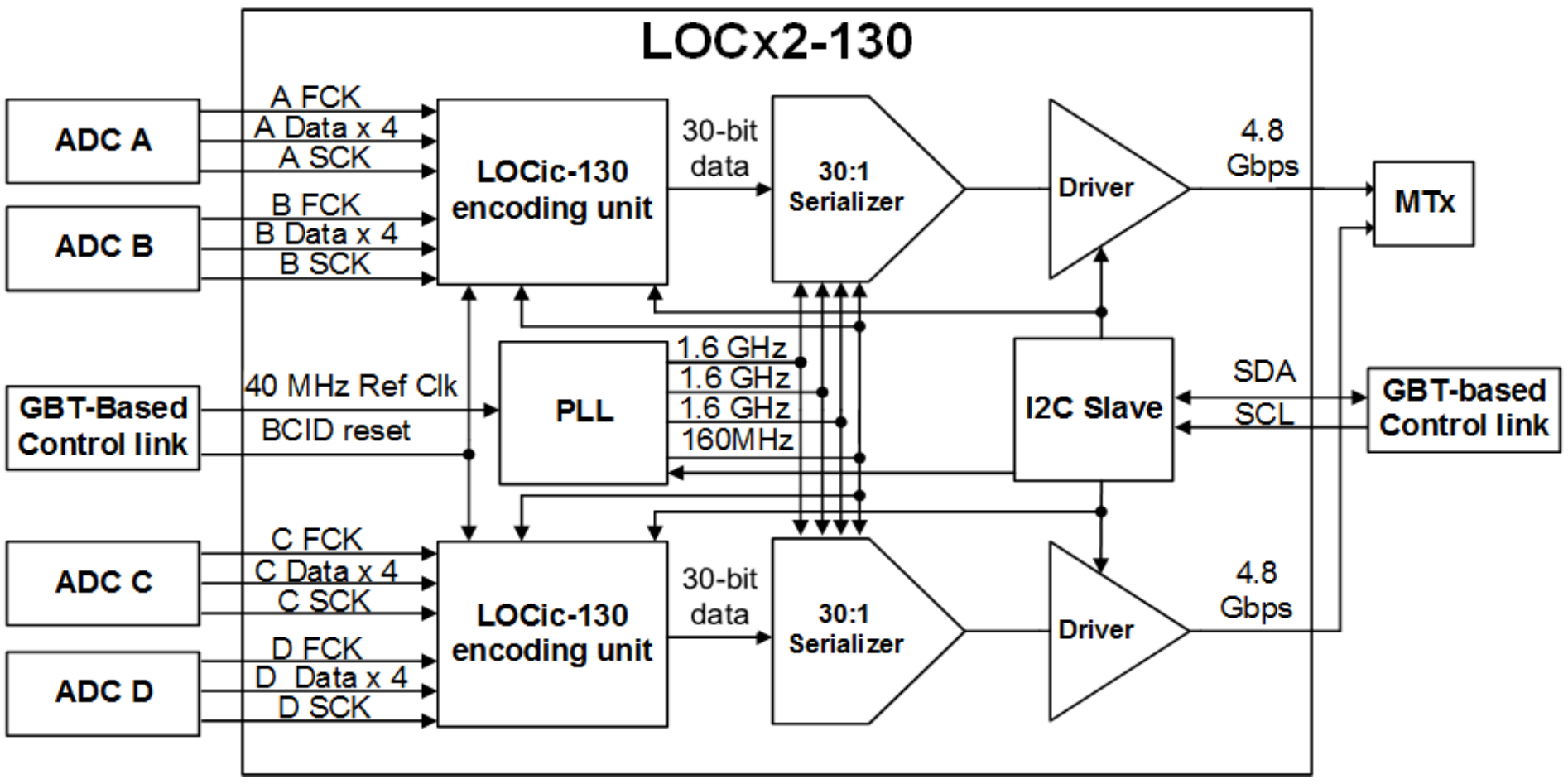

**Figure 3:** Block diagram of LOCx2-130.

LOCx2-130 is designed in a commercial 130-nm bulk CMOS technology and operates with a single power supply of 1.5 V. As a design goal, the silicon-proven components and the test facilities are reused as much as possible. In LOCx2-130, we employ a silicon-proven analog core of GBTX [15, 16] and TDS [17]. The analog core includes a PLL and a serializer. The block diagram of the serializer is shown in Figure 4. The PLL provides three 1.6 GHz clock signals ($Q_0$, $Q_1$, and $Q_2$) and a 160 MHz clock signal. The PLL has been carefully designed for radiation tolerance [15, 18]. Each serializer consists of 30-bit input registers, three 10-bit shift registers ($SR_0$, $SR_1$, and $SR_2$), and a 3:1 multiplexer. Shift-registers operate at 1.6 GHz. The timing diagram of the serializer is shown in Figure 5. The 1.6 GHz clock signals ($Q_0$, $Q_1$, and $Q_2$) have a duty cycle of one third and 120-degree phase difference from each other. The outputs of the three shift-register lines are multiplexed to assemble the full line rate. A detailed description of the serializer can be found in [15, 17].

We modify the analog core so that two serializers share a single PLL. This modification significantly reduces the power consumption. Since the analog core has been discussed elsewhere before [16, 17], we focus on the design of the encoding unit LOCic-130 in this paper.

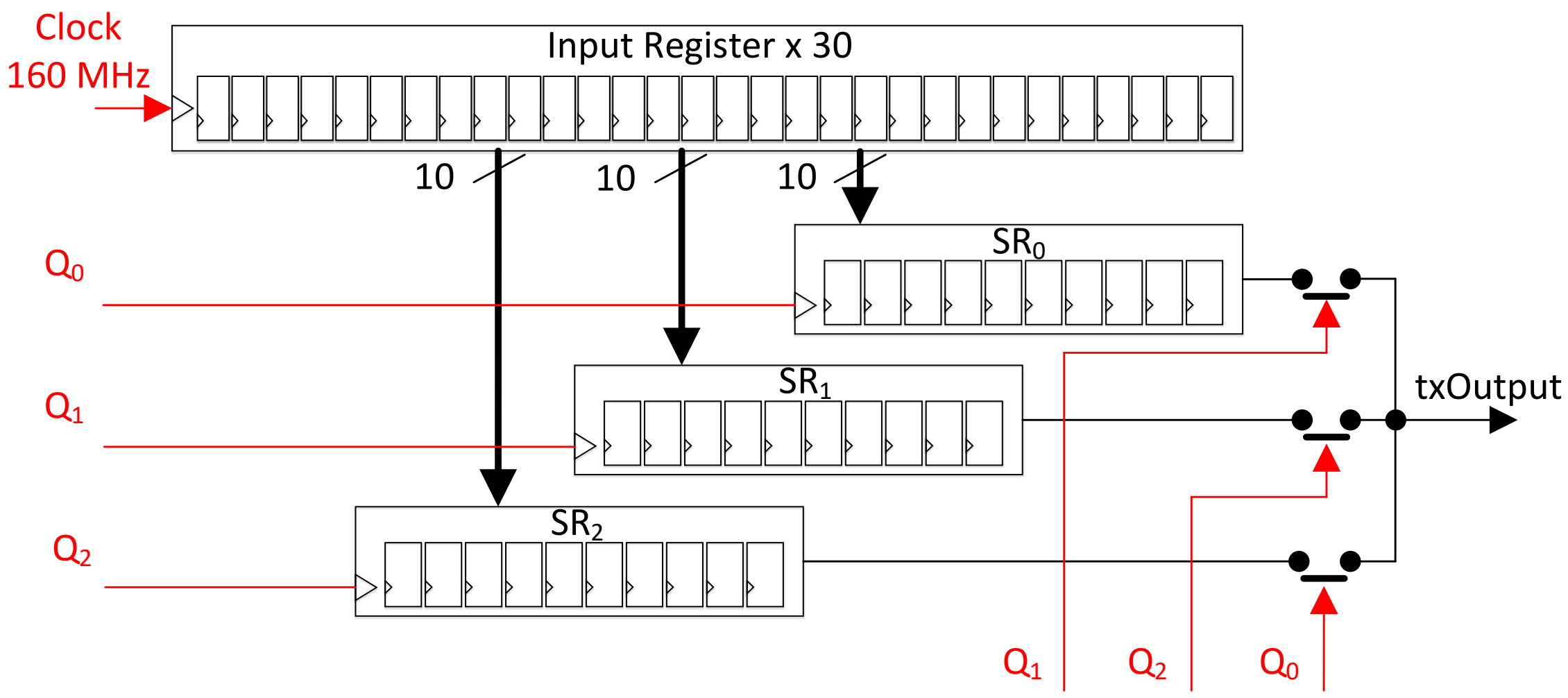

**Figure 4:** Block diagram of the serializer.

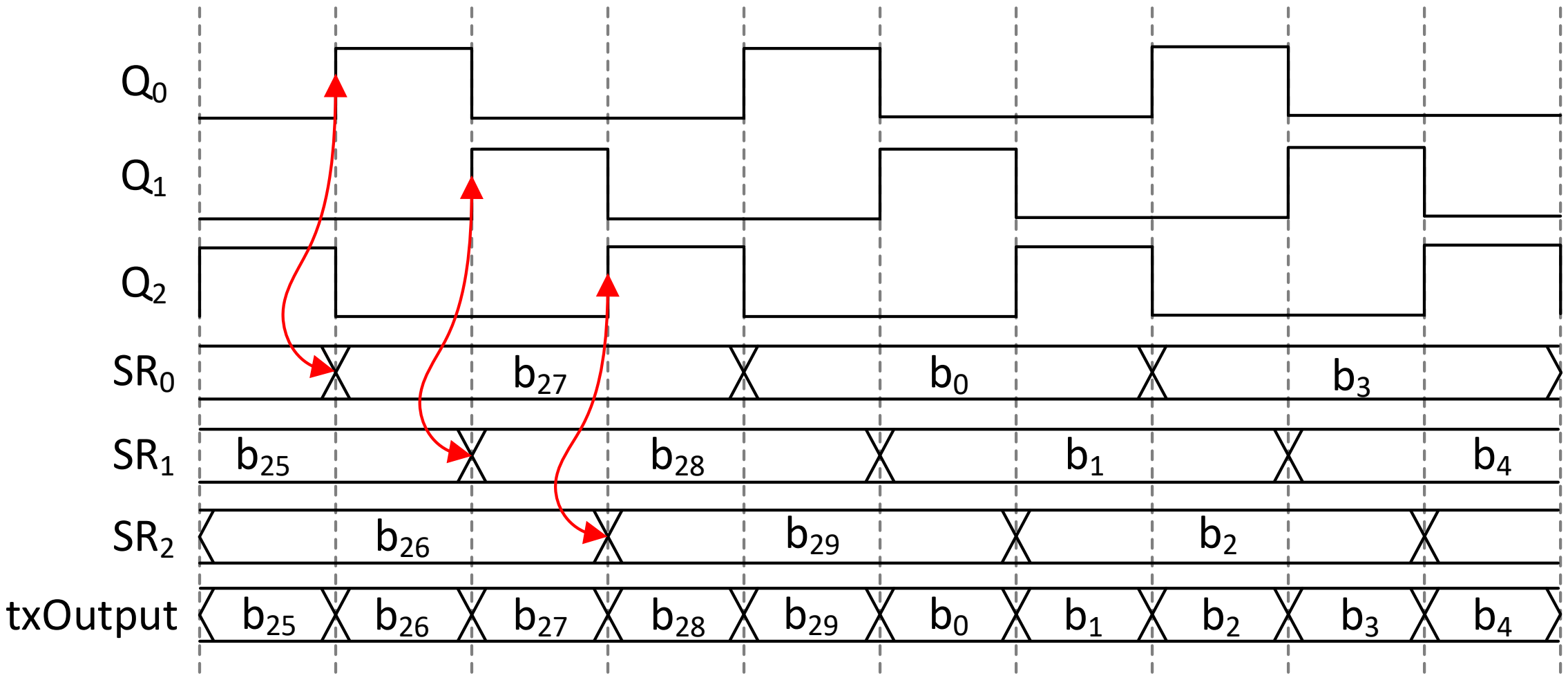

**Figure 5:** Timing diagram of the serializer.

The block diagram of the encoding unit LOCic-130 is shown in Figure 6. In addition to the core encoder, the LOCic-130 encoding unit also has an ADC interface module and a synchronous First-In-First-Out (FIFO) module. The ADC interface receives data from ADS5272, ADS5294, or Nevis ADCs and provides unified outputs for the FIFO. The FIFO is used to accommodate the different data rates of three types of ADCs. The core encoder is responsible for building the data frame and outputting to the serializer.

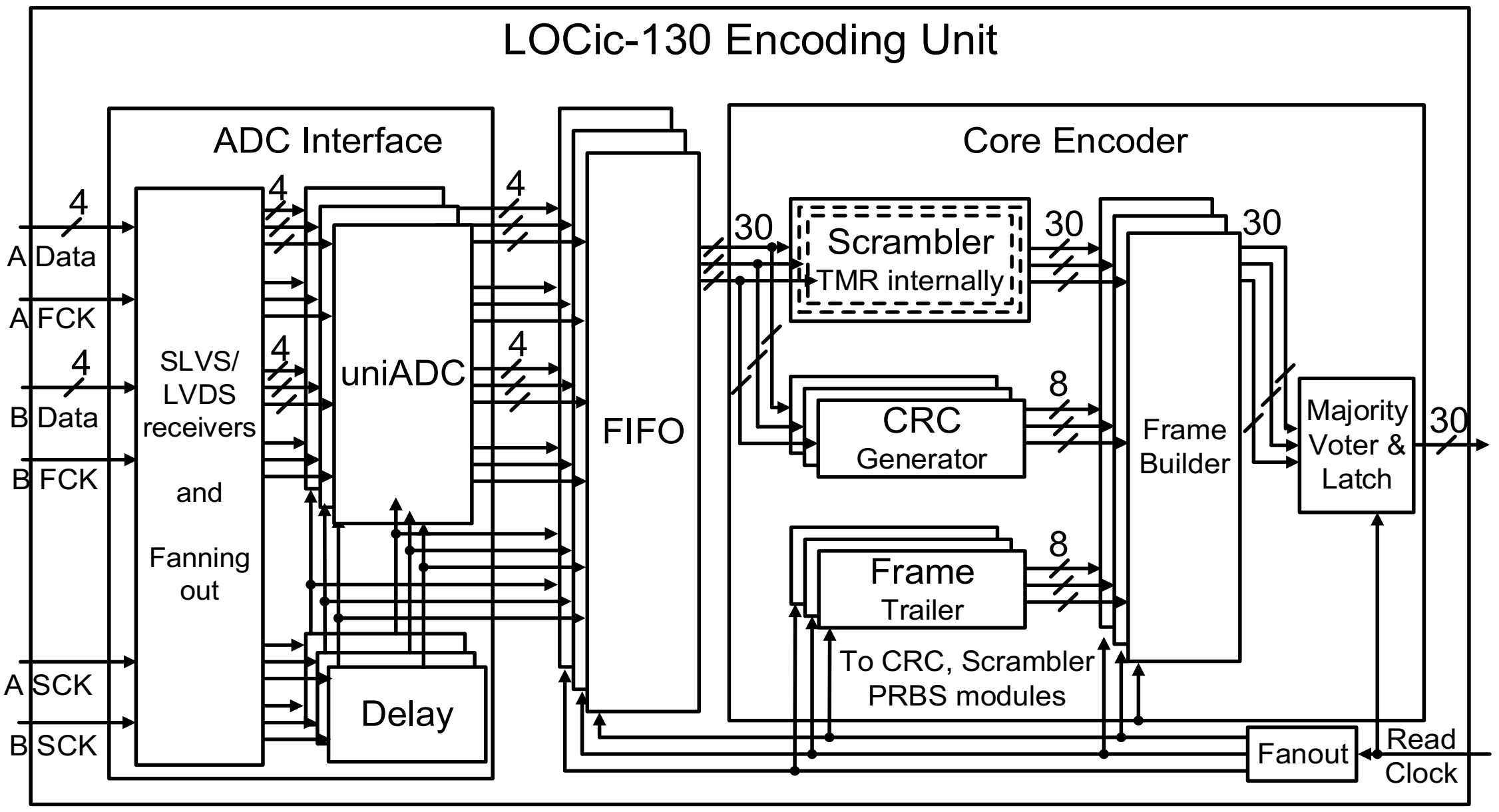

**Figure 6:** Block diagram of the LOCic-130 encoding unit.

### 3.1 ADC interface

The ADC interface is composed of Scalable Low-Voltage Signaling (SLVS) / Low-voltage differential signaling (LVDS) receivers, programmable delay cells, and universal ADC (uniADC) modules. The output signals of Nevis ADCs are in SLVS format, whereas the output signals of ADS5272 and ADS5294 are in LVDS format. The input SLVS/LVDS receivers accept both SLVS and LVDS signals. A programmable delay unit adjusts the delay of input SCK signals to ensure that uniADC acquires the ADC data and the FCK signals correctly. An optimal delay value, which is fixed once the system is set up, is determined at a system level after all values are scanned. When the core encoder works with a single ADS5272 or ADS5294 ADCs, uniADC makes a copy of the SCK and the FCK, so that the following circuits interfaces with two Nevis-like ADCs. The uniADC is also responsible for shifting the FCK signal to align with the first bit of the valid data. In the case of NEVIS ADCs and the calibration mode, the FCK signal is delayed one clock period of SCK to align with $D_{13}$. In the case of NEVIS ADCs and the data mode, the FCK signal is delayed two clock periods of SCK to align with $D_{11}$. In the case of ADS5272 and ADS5294, the FCK, which is aligned with $D_{11}$ and $D_{14}$, respectively, is not shifted. All the ADC data are provided to the following FIFO module, where the invalid data are dumped. Uniformly aligning the FCK signal with the first bit of the valid data for all cases simplifies the design of the FIFO.

### 3.2 FIFO

The FIFO is the interface circuit between the ADCs and the core encoder. On the input side, two Nevis-like ADCs write data into the FIFO in two independent clock domains. In the case of ADS5272 or ADS5294, the two Nevis-like ADCs have two of the same 240 MHz or 280 MHz FIFO write clocks, respectively. In the case of Nevis ADCs, two ADCs have two synchronous 320 MHz data clocks with independent phases. On the output side, the following core encoder reads the 30-bit data from the FIFO in a 160 MHz read clock domain.

The block diagram of the synchronous FIFO is shown in Figure 7. The FIFO has two independent write controllers, a read controller, and two memory units. The memory units are implemented directly using statistic D flip-flops (DFFs).

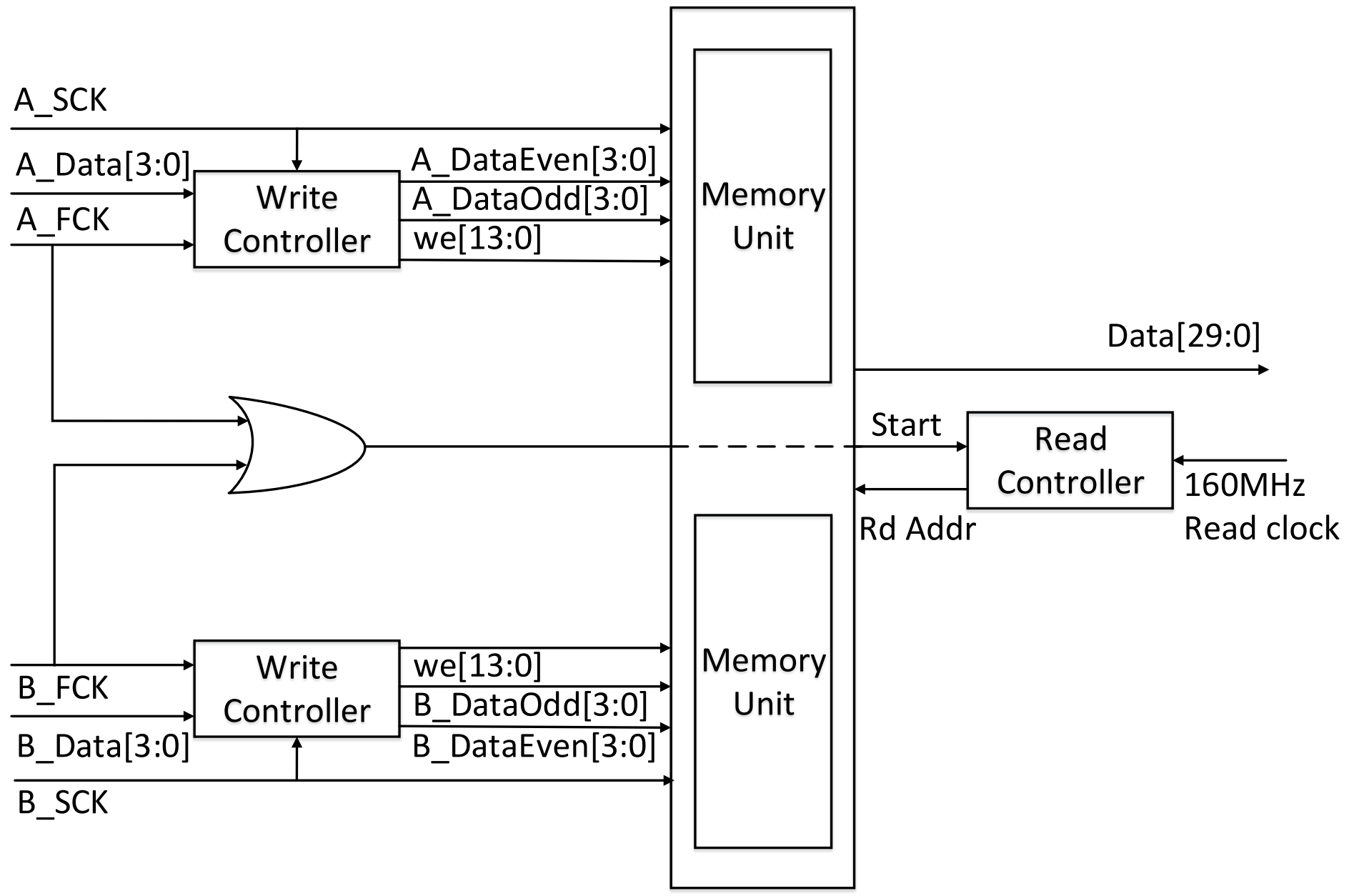

**Figure 7:** Block Diagram of synchronous FIFO

The two memory units have 112 cells in total and can store 2 (two memory units, one for ADC A and the other one for ADC B) × 14 (depth) × 4 (width) bits. Each write controller processes 4-channel ADC data. A simulated timing diagram of the write controller for ADC A is shown in Figure 8. To facilitate observation, the 4-channel ADC data A_Data[3:0] are simulated using linear values from 0 to 14. The input data are in a Double Data Rate (DDR) mode. The input data are latched at the rising edge and the falling edge of SCK to acquire the even-bit data ($D_{14}$, $D_{12}$, $D_{10}$, …, $D_0$) and odd-bit data ($D_{15}$, $D_{13}$, $D_{11}$, …, $D_1$), respectively. After the DDR process, A_DataEven[3:0] and A_DataOdd[3:0] are output to corresponding memory cells. The write controller is responsible for generating Write Enable signals (we[13:0]) for corresponding memory cells. Among the Write Enable signals, we[0], we[2], we[4], we[6], we[8], we[10], and we[12] are responsible for the even-bit data A_DataEven[3:0], whereas we[1], we[3], we[5], we[7], we[9], we[11], and we[13] are responsible for the odd-bit data A_DataOdd[3:0]. After an FCK arrives, the first bits of A_Data[3:0], which are aligned to the FCK signal in uniADC, of the valid data are written to the cells controlled by we[0], the second bits are written to the cells controlled by we[1], the third bits are written to the cells controlled by we[2], and so on. In the case of NEVIS ADCs and the calibration mode, after the bits $D_{13}$-$D_0$ are written, a new round of write-address-enable signals are generated based on the following FCK signal. Note that the bits $D_{15}$-$D_{14}$ are dumped. In the case of NEVIS ADCs and the data mode, the data $D_{11}$-$D_0$ are written to Cells 0-11, whereas the data $D_{15}$-$D_{12}$ are dumped. In the cases of ADS5272 and ADS5294, after $D_{11}$-$D_0$ and $D_{13}$-$D_0$ are written to the cells controlled by

we[0:11] or we[0:13], respectively, a new round of write-address-enable signals begins and no data are dumped.

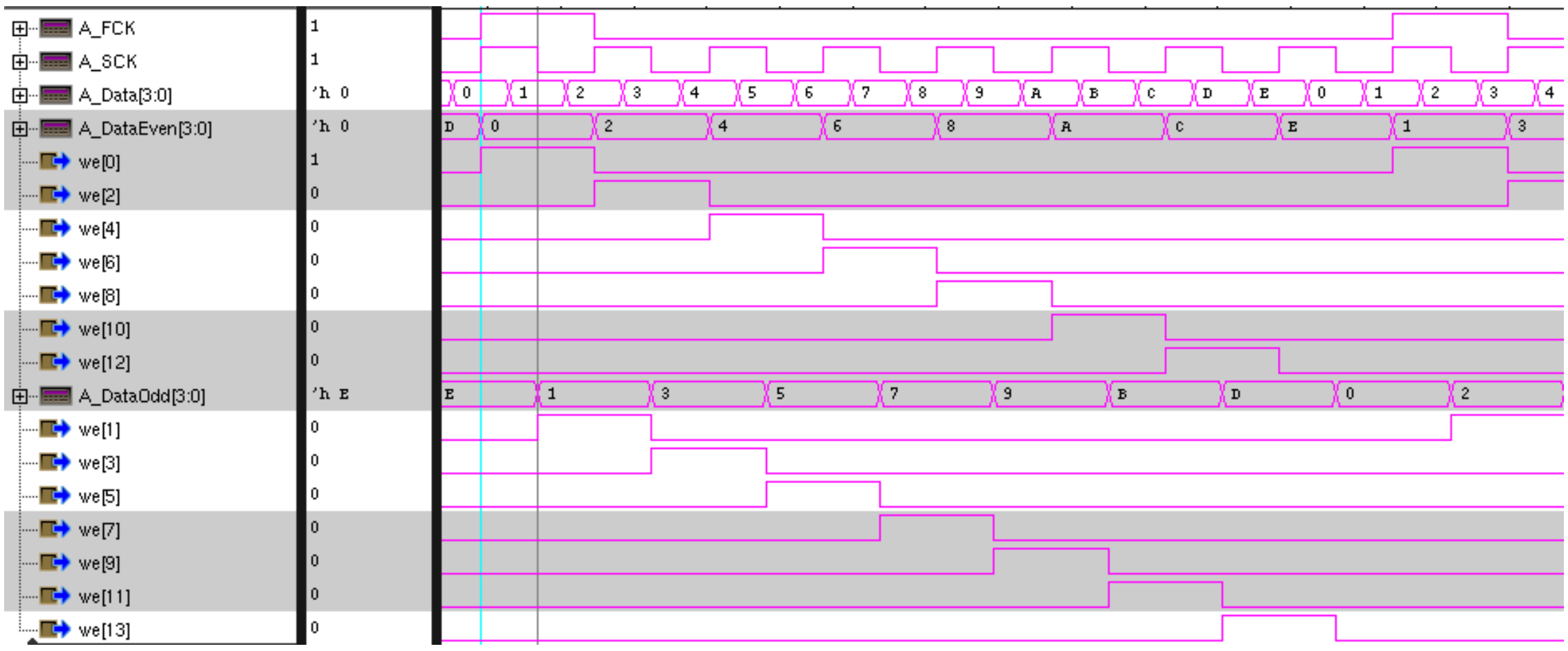

**Figure 8:** Timing Diagram of writer controller for NEVIS ADC A

The FIFO aligns the data from two Nevis ADCs with a timing skew up to 3.125 ns. Due to the skew between the two ADCs, the two independent write controllers may write one memory earlier than the other. The read controller controls the read operations of two memory units; hence the data from the two ADCs are aligned when they are read out. The two FCK signals start not only the round of write-enable signal sequences of the corresponding write controller but also the read address sequence of the read controller. As can be seen in Figure 7, the two FCK signals input an OR gate. The rising edge of the OR'ed FCK signal starts the read operation after a clock period of SCK. We adjust the time of the read operation to ensure that the read operation occurs after the data of the later ADC are written to the memory. Our simulation shows that in a 25-ns frame, the read operation does not catch up with the write operation, and the write operation of the next frame does not catch up with the preceding reading operation, either. Due to the limited memory depth, it is required that the skew between ADC A and ADC B are no more than a clock cycle of SCLK, i.e., 3.125 ns.

The write operation is always in the DDR mode. The read clock is 160 MHz and the read data bus width is 30 bits. Each data frame needs 4 read operations in single data rate mode to read out the 112 bits of data and extra 8 bits of zero from the FIFO. Thus, the different data rates for the three types of ADCs are accommodated in the FIFO. In the data mode, there are only 6 bits of valid ADC data in the last read operation, whereas in the calibration mode, there are 22 bits of valid ADC data in the last read operation. The following core encoder replaces the invalid data of the received 120 bits of data. In the calibration mode, an 8-bit trailer is used to replace the last 8 bits. In the data mode, the 16-bit CRC and the 8-bit trailer are used to replace the last 24 bits.

### 3.3 Core encoder

The core encoder builds a data frame and outputs to the serializer in 30-bit width at 160 MHz. The core encoder includes a scrambler, a CRC generator, a frame trailer generator, and a frame builder.

The scrambler generates balanced codes without extra cost. The scrambling polynomial is $x^{58}+x^{39}+1$, which is widely is used in industry. In the implementation, both the input and output of the scrambler are 30-bit parallel data to match the 30:1 serializer. The basic logic operation of the scrambler is the exclusive-or (XOR) operation. In each clock cycle of 160 MHz, the scrambler inputs a 30-bit payload and produces 30-bit scrambled data, using a parallel algorithm we derived.

The CRC module generates 16-bit CRC code to detect if a potential bit-flip error has occurred in the data transmission. We use the polynomial $x^{16}+x^{14}+x^{12}+x^{11}+x^{9}+x^{8}+x^{7}+x^{4}+x+1$ to calculate the CRC from the unscrambled payload. Our studies show that a single bit flip in the payload may end up with 1, 2, or 3 bit flips in the receiving end after descrambling. In fact, the descrambling process always amplifies a single bit flip into 3, but the three bit flips may cross a frame boundary, resulting in 1, 2, or 3 bit flips in one frame. If a burst error occurs in the data transmission, the bit-flip error may last tens of bits. All odd bit flips or up to 4 even bit flips can be detected by our CRC code. For other even bit flips, the probability that CRC code misses the error is about 0.003%. In the parallel CRC algorithm, in each clock cycle, a CRC code of 30-bit data is calculated. The whole 96-bit data takes 4 clock cycles to process. The CRC process is in parallel with scrambling and has no extra latency penalty.

The frame trailer generator is responsible for producing a fixed 1010 pattern and the BCID field. The BCID field takes two bits from a PRBS $2^7$-1 generator and two bits from a PRBS $2^5$-1 generator. A BCID reset signal resets both PRBS generators. The frame builder assembles the data into 30-bit width data and feeds it to the 30:1 serializer.

### 3.4 Triple Modular Redundancy (TMR) design

The Verilog code of LOCic-130 is triplicated. The whole layout and the pin map of the triplicated LOCic-130 have been generated through the digital design flow. The function and timing of the design have been verified at the post-synthesis level. The diagram of LOCic-130 with TMR structure is shown in Figure 6. The modules that have pipeline structure without any internal feedback, like the FIFO and the Frame Builder, are simply instantiated three times for triple redundancy. The CRC generator and the Frame Trailer, which both have internal feedbacks but are reset periodically, are also simply triplicated. It is best to add a majority voter in every block to eliminate potential parallel SEU events in different branches. However, due to the tight timing margin of the design, no majority voter is implemented in these blocks and all SEU events will be flushed out in the next clock cycle. The Scrambler, however, which has a feedback and no reset mechanism, is triplicated internally with the voter added at the input of every DFF to eliminate errors so that an error will not be latched in the DFF and does not cause permanent malfunction. After the triplicated Frame Builder, a final Majority Voter & Latch is added. The outputs of LOCic-130 with TMR are the 30-bit voted parallel data.

### 3.5 Layout

The layout of LOCx2-130 is shown in Figure 9. The die area of LOCx2-130 is 2.0 mm × 5.0 mm. The analog core, including the PLL and the two 30:1 serializers, is located at the right of the floor plan and has larger decoupling capacitors, the central pink area in the figure, than the other circuits. The digital functional blocks, including the SLVS receivers (located on the top and bottom edges), two LOCic-130 encoding units, and the I2C slave, are noisier than the analog core. We carefully isolated the substrates of these circuits and provide separated power supply and ground for each of them.

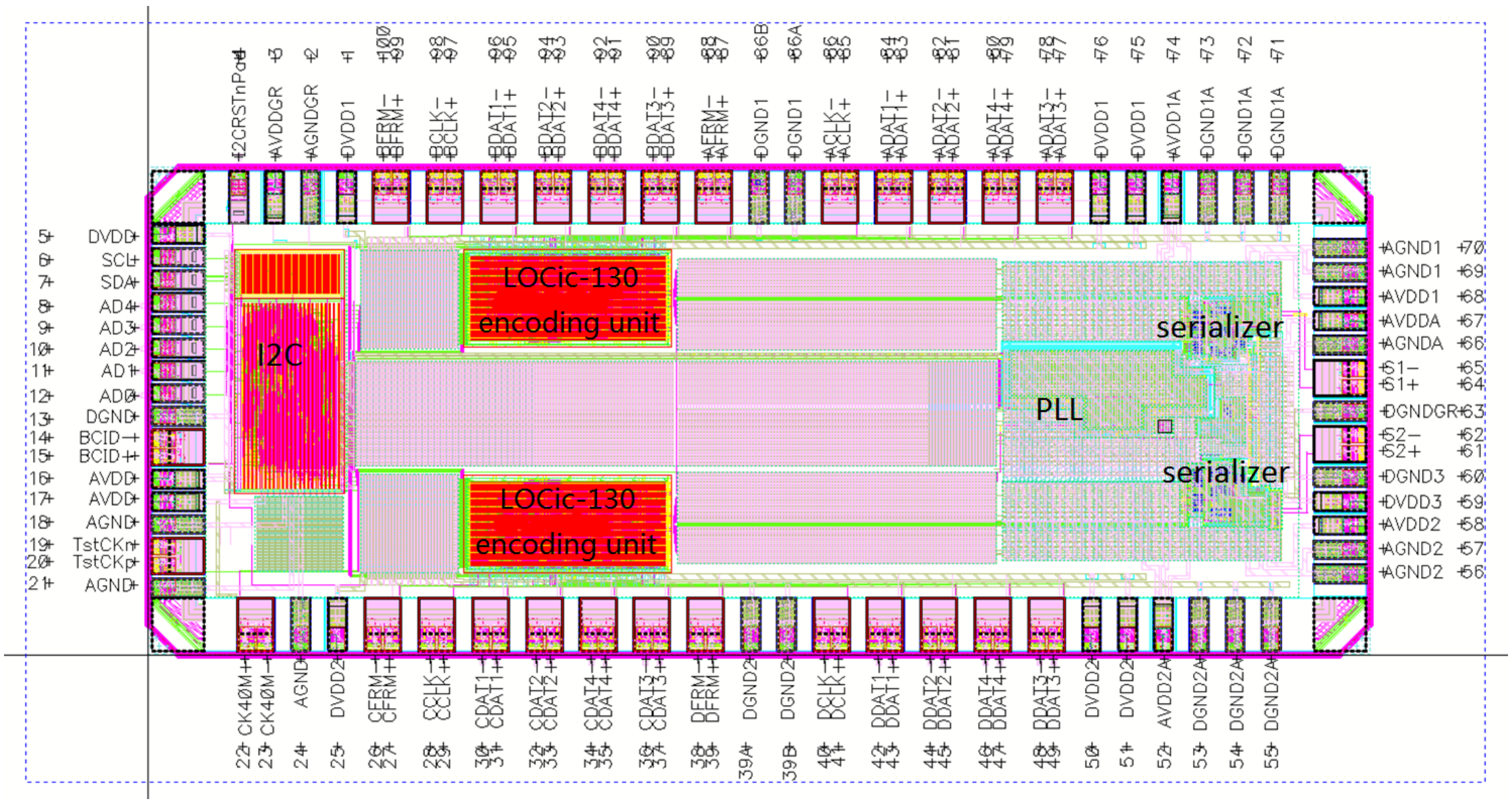

**Figure 9:** Layout of LOCx2-130.

### 3.6 Package

LOCx2-130 is packaged in a 100-pin plastic quad-flat no-leads (QFN) package. Figure 10 is a picture of two packaged chips.

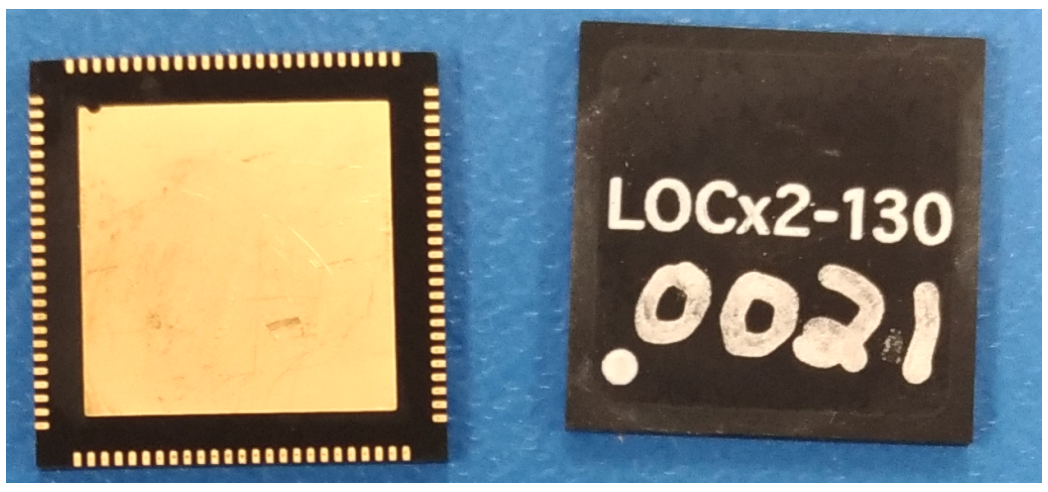

**Figure 10:** Bottom view (left) and top view (right) of QFN packaged ASIC.

## 4. Characterization of LOCx2-130

The performance of the LOCx2-130 prototype has been characterized. The performances include eye diagrams, bit error rate, latency test, and radiation tolerance.

### 4.1 Test setup

The block diagram of the test setup is shown in Figure 11. During the test, a Xilinx Kintex-7 FPGA emulated ADC and a BCID reset generator. The emulated ADC data were aligned with FPGA-embedded I/O delay modules. The FPGA also implemented two link receivers and an error logger. A clock board generated all clocks used in the test. The serial outputs of LOCx2-130 were sent either back to the FPGA to measure bit error rate or to a high-speed real-time oscilloscope (Model DSA 72004 produced by Tektronix) to measure eye diagrams and jitter. An $I^2C$ master configured LOCx2-130 in the test. A picture of the test setup is shown in Figure 12.

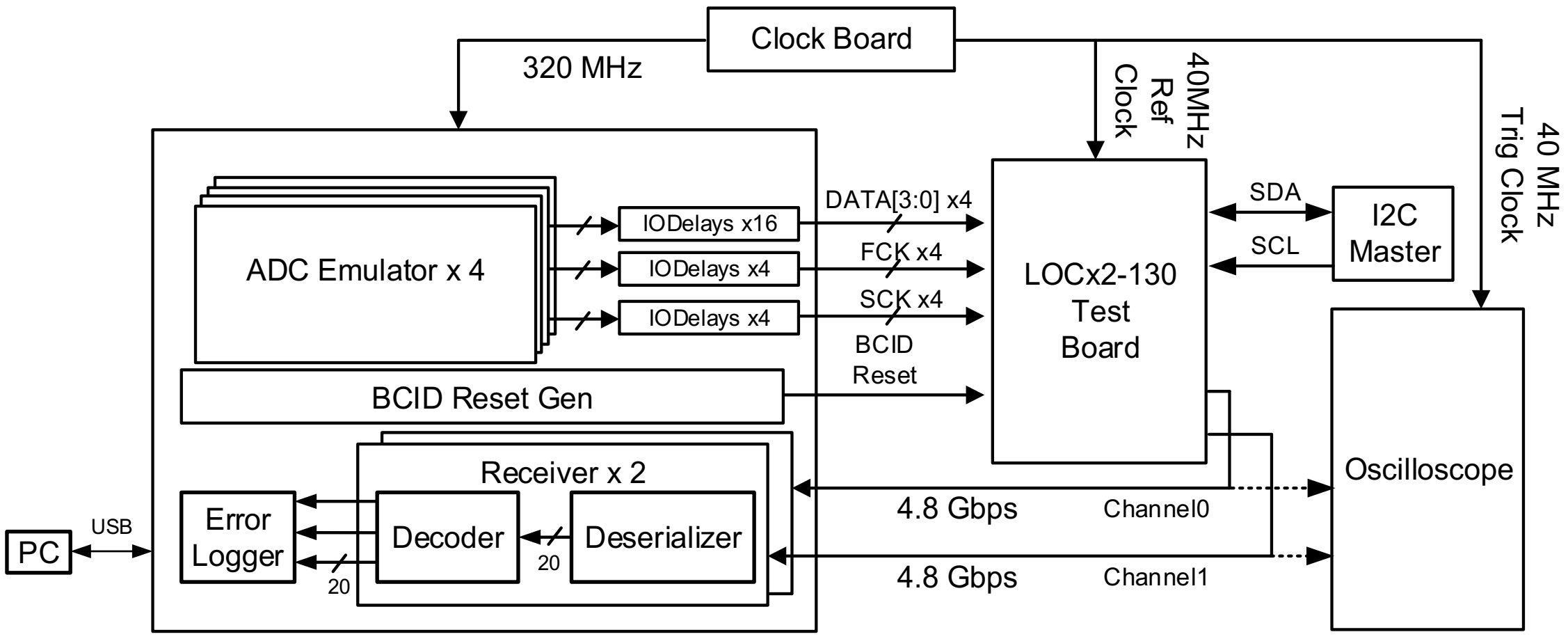

**Figure 11:** Block diagram of the test setup.

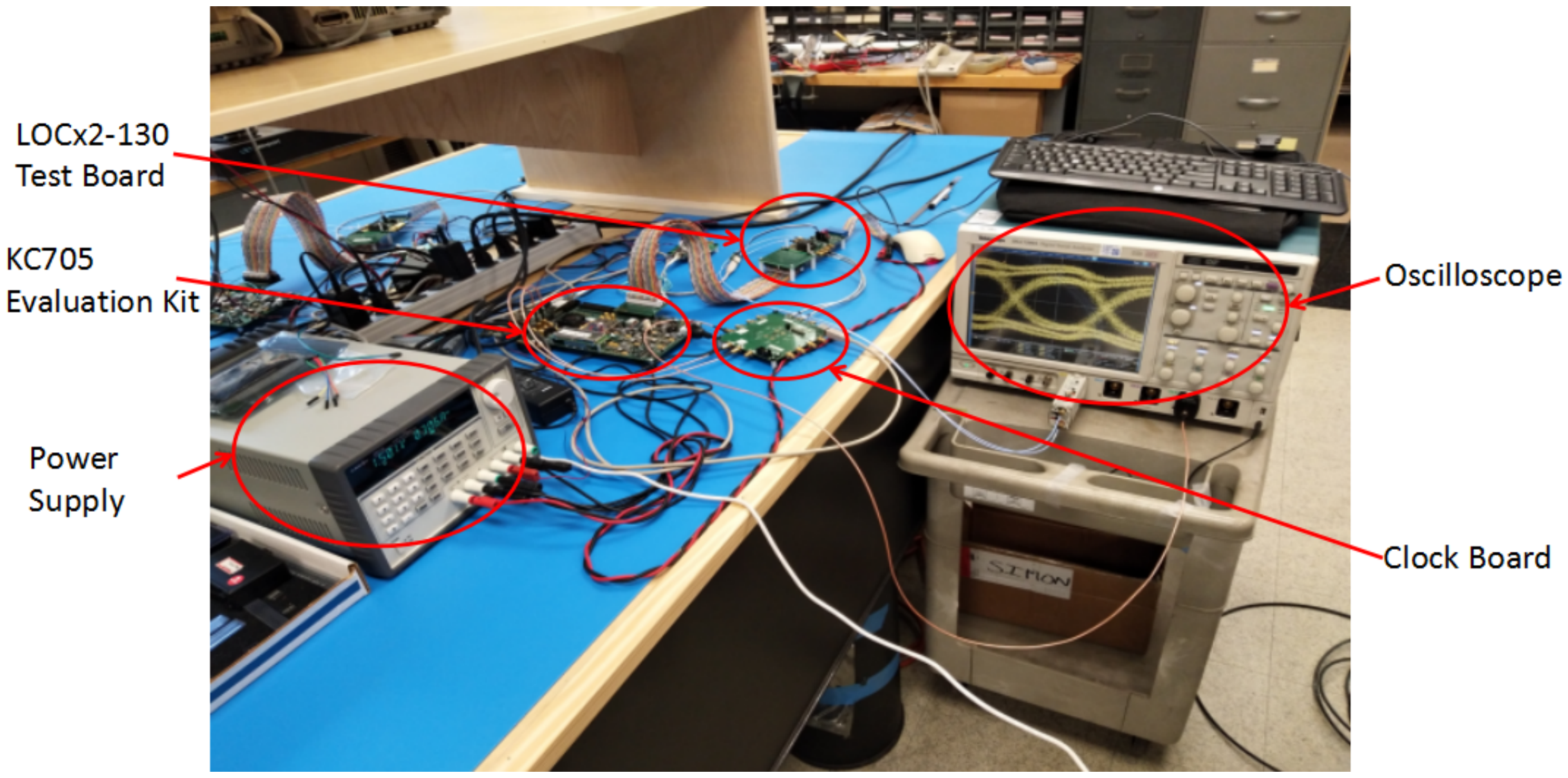

**Figure 12:** Picture of the test setup.

### 4.2 Eye diagram test

Figure 13 shows an eye diagram of LOCx2-130 at 4.8 Gbps. Both rise time and fall time are about 78 ps. Random jitter is 2.3 ps (RMS) and deterministic jitter is 28 ps (peak-peak). Correspondingly, total jitter is 52 ps (peak-peak) at the bit error rate of $10^{-12}$. The typical output amplitude is 300 mV (peak-peak).

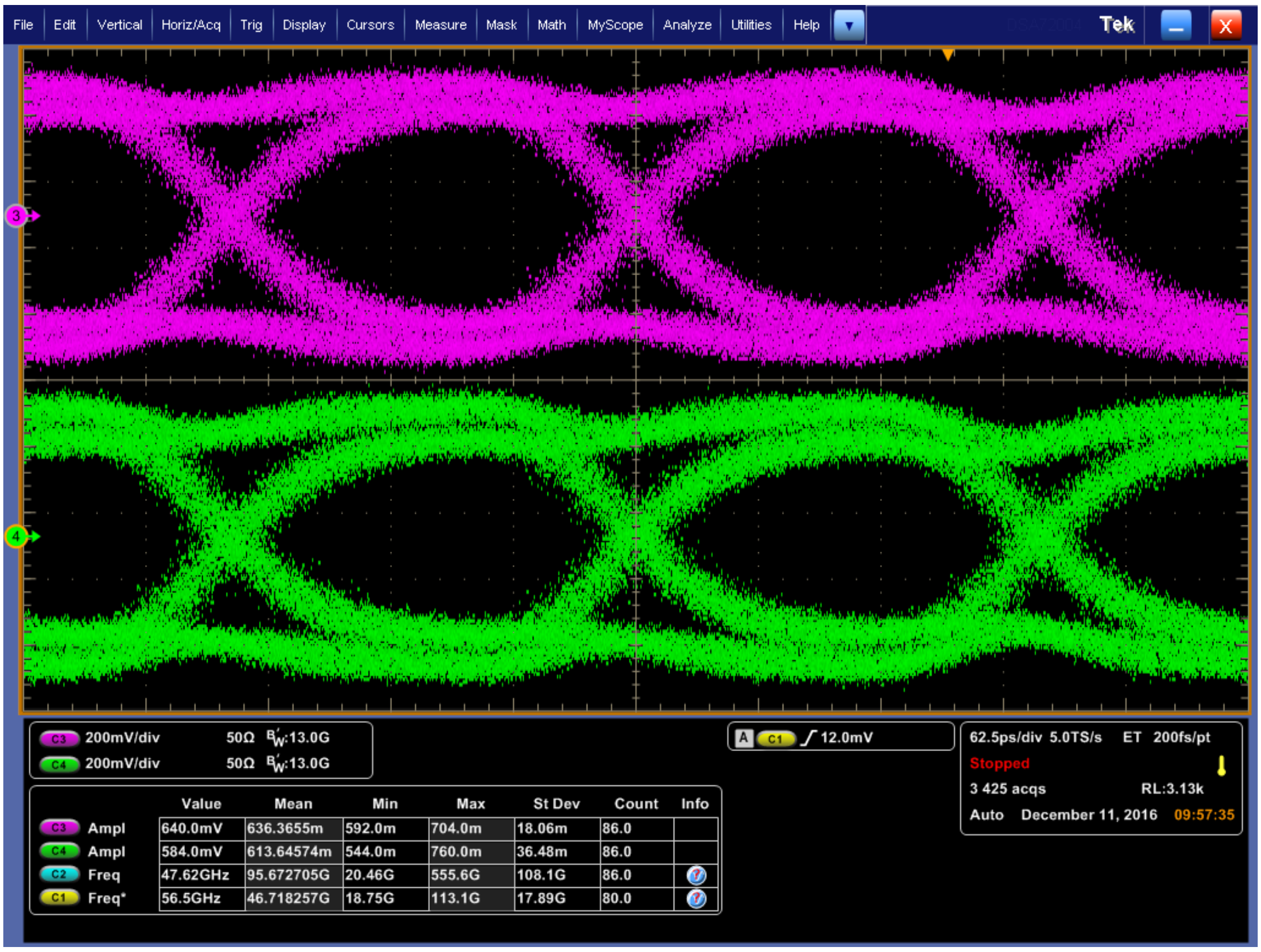

**Figure 13:** Eye diagram of LOCx2-130 at 4.8 Gbps.

### 4.3 Bit error rate test and latency measurement

The bit error rate of LOCx2-130 was tested in the lab. No error was observed in a nine-minute test. Thus, the corresponding bit error rate can be calculated to be less than $10^{-12}$ at the confidence level of 95%.

The measured power consumption of LOCx2-130 is about 440 mW at room temperature and 4.8 Gbps. The power consumption of LOCx2-130 is about half of that of LOCx2. The corresponding power efficiency is 45.8 mW/Gbps, less than half of the design goal of 100-mW/Gbps. The primary reason to achieve such power efficiency is the advanced technology. The power voltage of LOCx2-130 is 1.5-V, whereas that of LOCx2 is 2.5 V. The secondary reason is that the core encoder of LOCx2-130 operates at 160 MHz, whereas that of LOCx2 operates at 320 MHz.

Table 1: Latency summary

| Function block | | Latency (ns) |
|---|---|---|
| TX | FIFO | 16.4~22.7 |
| | Scrambler & CRC generator & Frame builder | 6.3 |
| | Serializer | 11.7 |
| | Total | 34.4~40.7 |
| RX | Deserializer | 36.7~40.1 |
| | Data Extractor | 12.5 |
| | Descrambler | 4.2 |
| | CRC Checker | 4.2 |
| | Total | 57.6~61.0 |
| **Total** | | 92.0~101.7 |

We measured the latency of the whole link, which includes LOCx2-130 and the corresponding receiver but excludes the optical transceiver and the optical fiber. The measurement results of the latency are shown in Table 1. The latency of the whole link is from 91.9 ns to 101.7 ns and that of LOCx2-130 is from 34.4 ns to 40.7 ns, achieving the design goal.

### 4.4 Irradiation tests

Two prototype chips of LOCx2-130 were irradiated in x-rays with the maximum energy of 160 keV and the peak energy of about 30 keV after a 2-mm aluminum filter. The total ionizing dose reached 3.0 kGy, exceeding the requirement of the ATLAS LAr Calorimeter Phase-I trigger upgrade. Eye diagrams and power consumption were measured before and after irradiation. No significant changes in power consumption and eye diagrams were observed.

Two prototype chips were tested in a proton beam. The beam energy was chosen to be 200 MeV, per the ATLAS radiation policy on radiation tolerant electronics [7, 19]. The beam diameter was about 1 cm. The test setup was similar to the bit error rate test described above. Only the LOCx2-130 being tested was exposed in the beam, whereas all other circuits were shielded, about 0.3 meter off the beam. A personal computer ran in the control room, which was about 15 meters away from the beam. A picture of the test setup is shown in Figure 14. The beam was coming down from the top. The accumulated fluences in the test were $9.2\times10^{9}$ and $8.7\times10^{9}$, respectively. Neither bit-flip error nor burst error was observed in any device being tested. The bit error rate in the future HL-LHC application was extrapolated to be less than $8.5\times10^{-16}$ at the confidence level of 95%. Since the same PLL design is employed in LOCx2-130 as in GBTX [19], due to the limited fluences accumulated in the test, the burst error rate is not extrapolated based on our test results.

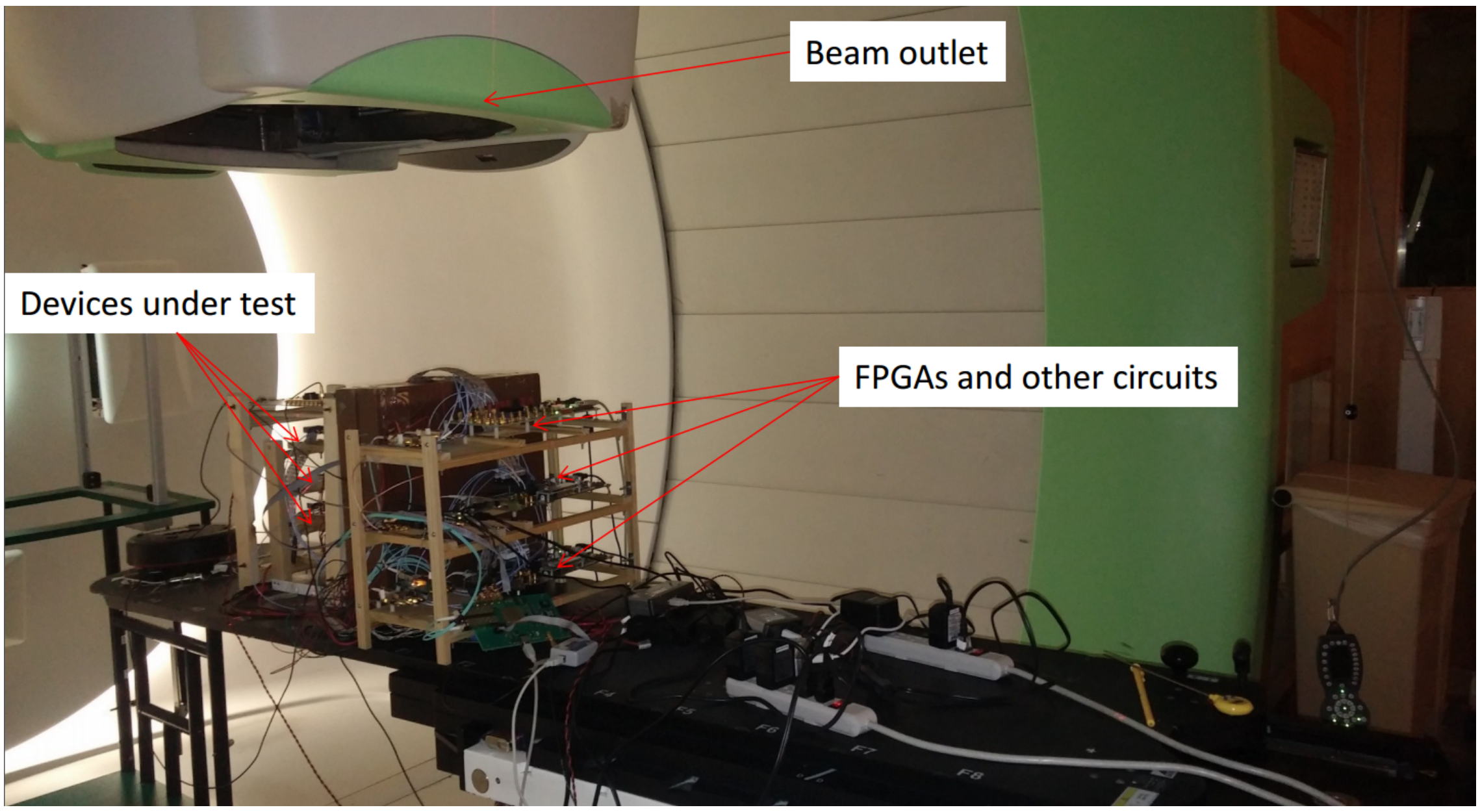

**Figure 14:** Picture of proton test setup.

## 5. Conclusion

The design and test results of a transmitter ASIC, LOCx2-130, are presented. LOCx2-130 is a two-channel transmitter designed for the ATLAS LAr Calorimeter Phase-I trigger upgrade. Each channel of LOCx2-130 encodes ADC data and serializes data at 4.8 Gbps with a latency of less than 40.7 ns. The power consumption of LOCx2-130 is about 440 mW.

## Acknowledgments

We acknowledge the support by the NSF and the DOE Office of Science, SMU's Dedman Dean's Research Council Grant, and the National Natural Science Foundation of China under Grant No. 11705065. We are grateful to Dr. Paulo Moreira, Syzmon Kulis and Sandro Bonacini from CERN, Drs. Jinhong Wang and Junjie Zhu from University of Michigan in Anna Arbor for their assistance in the ASIC design. We thank Drs. Mitch Newcomer and Dong Lei from University of Pennsylvania for their help in the irradiation test. We thank Drs. Hucheng Chen, Kai Chen, and Hao Xu from Brookhaven National Laboratory, Dr. Nicolas Dumont Dayot from LAPP, and Dr. Bernard Dinkespiler from CPPM for beneficial discussions during the design and testing process.